# A nanoindentation study of attenuation in geological materials

Nir Z. Badt, Ron Maor, and David L. Goldsby

University of Pennsylvania

June 13, 2024

Key points

- A new method for measuring the attenuation spectrum of geological materials by nanoindentation at room temperature.
- PMMA and indium attenuation spectrum from nanoindentation are in excellent agreement with results obtained by other methods.
- Olivine and quartz attenuation at room temperature is measurable frequencies < 0.001 Hz.

**Abstract**

Viscoelastic behavior in geological materials controls a wide range of geophysical phenomena, such as mantle convection. We present a new method for measuring attenuation in single crystals of minerals and in reference materials over a frequency range of $1-10^{-4}$ Hz via nanoindentation. In the experiments, we calculate the phase lag between a sinusoidal load applied to the nanoindenter's tip and its displacement into and out of the tested sample, which provides a measure of the inverse quality factor $Q^{-1}$ (i.e., attenuation) of the sample. Experiments were conducted on polymethyl methacrylate (PMMA), indium, halite, olivine and quartz. Attenuation spectrum from our tests on PMMA and indium are in excellent agreement with reported values from past studies. We quantified the natural damping of the nanoindenter and show that it becomes comparable to that of the samples at frequencies greater than 0.1 Hz, but is much lower at lower frequencies.

**Plain language summary**

Geological materials display behave as viscoelastic solids, exhibiting elastic and viscous behavior as they deform. Viscoelasticity is tied to numerous geophysical phenomena, such as post-glacial isostatic rearrangement. Quantifying the viscoelastic behavior of rocks and minerals through experimentation is important for constraining geodynamical models and for extrapolating these models to make useful predictions. Measuring the viscoelastic deformation of these materials is, however, challenging due to the experimental conditions that are required to activate these deformation mechanisms. We present a new method for measuring the time-dependent viscoelastic response of quartz, halite and olivine using nanoindentation. The new method is easy to set up and allows for testing with high spatial resolution. We apply a sinusoidal load on the indenter's diamond tip which indents into and out of the tested specimen. We measure the phase lag between the sinusoidal load and the sinusoidal displacement of the tip, which is the material's viscoelastic response (attenuation). To test the validity of our method we also measure the attenuation spectrum of indium and PMMA and compare these measurements to reported attenuation spectrums from other apparatuses. Our results are in excellent agreement with reported attenuation for indium and PMMA, showing the robustness of our method.

# Introduction

Understanding the viscoelastic behavior of geological materials is important for modeling numerous terrestrial and planetary phenomena, including seismic wave attenuation, tidal dissipation, interseismic surface deformation and postglacial isostatic rebound. Quantifying the viscoelastic behavior of geomaterials via laboratory experiments has been the focus of multiple studies in past decades (e.g., Anderson et al., 1996; Anderson et al., 2001; Wang et al., 2012; Cao et al., 2021; Qu et al., 2021) but the number of such studies has lagged observations tied to viscoelastic behavior by the growing use of remote-sensing methods (e.g., Khurana et al., 1998; Pappalardo et al., 1999; Hearn, 2003; Wang et al., 2020). In the case of geological materials that are found in the earth's lithosphere, this comparative lack of experimental data is due to the difficulty of measuring viscoelastic behavior at high temperatures and elevated confining pressures. Furthermore, the timescale of the viscoelastic response in the earth in slow processes like postglacial rebound operate over time periods of $10^{10}$ s (Karato and Spetzler, 1990), placing further constraints on experiments.

The viscoelastic response of geomaterials is typically measured in the laboratory by subjecting a test specimen to sinusoidal loading and measuring its sinusoidal strain response (e.g., Jackson, 1993; Tan et al., 2001; Cooper, 2002; Jackson et al., 2002; Chapman et al., 2019). For sub-resonant dynamic methods, a loss angle $\delta$ is equal to the phase angle $\phi$ between an applied sinusoidal force and the resulting sinusoidal displacement (or strain) of the specimen (Lakes, 1998). The forced oscillation method is commonly used in measurements of $Q^{-1}$ on rocks and minerals, where $Q^{-1} = \tan \phi$ (Lakes 1998; Cooper, 2002), as a means of characterizing the anelastic response of geomaterials as a function of the forcing frequency $f$ (Jackson, 1993; Cooper, 2002). We note here that some studies refer to the tangent of the phase angle $\tan \phi$ as `internal friction' (e.g., Jackson, 1993); to avoid confusion we refer to any strain energy dissipation during cyclical loading as attenuation $Q^{-1}$, the inverse of the quality factor $Q$. In order to activate a viscoelastic response and mitigate brittle fracturing in crystalline solids, such as rocks and minerals, cyclic loading is sometimes conducted at elevated confining pressures and temperatures. There are only a few apparatuses with such capabilities, such as a high-pressure gas apparatus (e.g., Jackson, 1993; Jackson et al., 2002) and the deformation-DIA (D-DIA) apparatus (Cao et al., 2021; Hein et al., 2022). The D-DIA is operated on an x-ray beamline, allowing the use of X-ray diffraction and radiography to measure in-situ stress and strain, respectively, during cyclic loading.

We employ a new method to measure attenuation in minerals and other materials using the forced oscillation method adapted for the nanoindenter. Advantages of conducting attenuation experiments in a nanoindenter are that 1) the experiments are easy to perform compared to those in a traditional high-pressure apparatus, and 2) the nanoindenter provides very precise control of the sinusoidal load and precise measurements of the resulting sinusoidal displacement. Our experiments were conducted on samples at room temperature only. However, our study may lay the foundation for conducting attenuation experiments at elevated temperatures using a dedicated sample heating stage. The large hydrostatic component of stress in the sample beneath the indenter tip provides confinement of the sample – in effect, the sample becomes its own pressure vessel – allowing one to study plastic deformation in minerals even at room temperature. Our method could eventually be used as a simple yet powerful method for measuring attenuation of rocks and minerals at the temperature and pressure conditions of the earth's interior.

## Methods

### Apparatus and sample preparation

Cyclical loading experiments were conducted using an iMicro nanoindenter (Nanomechanics Inc.) located at the University of Pennsylvania (Thom et al., 2018; Thom et al., 2019). The nanoindenter contains an electromagnetic coil that can apply up to 1 N of force to the diamond Berkovich indenter tip. Single crystals of the minerals San Carlos olivine, quartz, and halite, of unknown origin and crystallographic orientation, and samples of PMMA (plexiglass) and indium were also tested. Each specimen was mounted on an aluminum puck with a diameter of 25.4 mm with Crystalbond 509 adhesive, and the sample surfaces were polished with a suspension of $0.025-\mu m$ aluminum oxide in water on a felt-covered polishing wheel. Finally, the surface of each specimen was cleaned with ethanol just before each experiment. All experiments were conducted at room temperature and at a relative humidity of ~25%.

### Forced oscillation method and phase-angle calculation

We employed the forced oscillation method (Lakes, 1998) for measuring the phase-angle $\phi$ between the applied force and the displacement (or depth $h$) of the indenter tip into and out of the sample surface (figure 1). The experimental procedure was identical for all five materials: (1) the indenter tip was lowered until it contacted the surface of the specimen; (2) the tip was loaded at a constant rate of $\dot{P}/P$ = 0.2 s$^{-1}$ to a final load of $P$ (specified in figure 3); (3) cyclical loading was initiated following a force oscillation function ($P_{test}$) of the form $P_{test} = P + A \sin 2\pi f$ (figure 1a), where $A = 0.1P$ was the amplitude of the oscillations and $f$ their frequency (where the oscillation period $T = 1/f$). Cyclical loading was conducted for a duration of ten cycles for $T \leq 1000$ s and five to eight cycles for $T > 1000$ s. At least two tests were conducted at each frequency for each specimen to check for reproducibility and the robustness of the testing method, except for one test in quartz with $T = 10^4$ s.

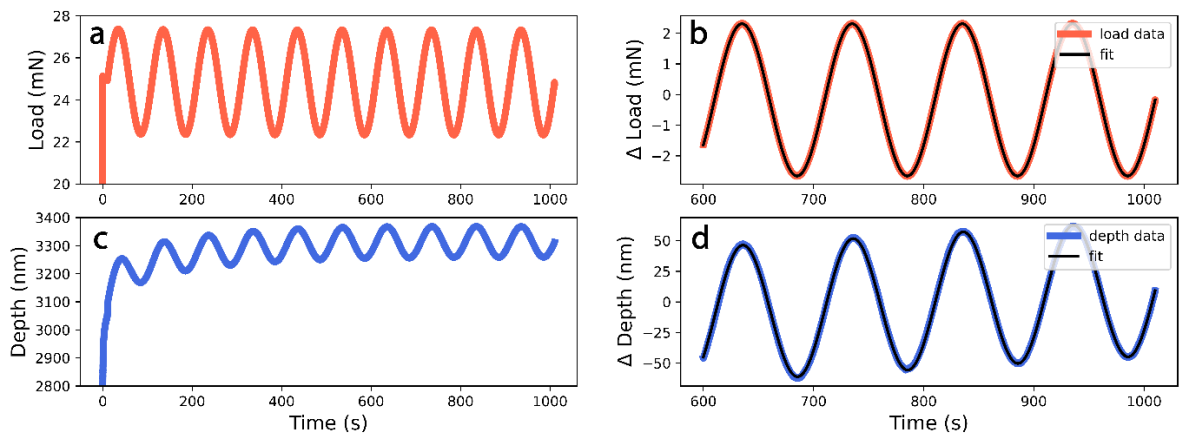

Figure 1. Forced oscillation test on PMMA with oscillation period of 100 s (a) and the subsequent change in indenter depth with time (c). Fits to the load (b) and depth (d) are presented for the last four oscillations in the test, and the mean is deducted from the data and the fits for both (b) and (d).

The phase angle $\phi$ between the load and the depth data was calculated by fitting an anonymous sine function $F$ to the load and depth data over the last four oscillations (or the last three oscillations where $T \geq 1000$ s) with a least-squares fit (McCarthy et al., 2011; Cao et al., 2021). $F$ was defined as:

$$F = a_{est} \sin(2\pi f t + \phi_{est}) + m_{est}, \qquad (1)$$

where the estimated parameters are the amplitude $a_{est}$, phase $\phi_{est}$ and mean $m_{est}$, and $f$ is the prescribed oscillation frequency (known for each test) and the time series $t$. $F$ was fit to the load (figure 1b) and depth (figure 1d) with the estimated phase for load $\phi_p$ and for depth $\phi_d$, where the coefficient of determination for the fitting algorithm $R^2$ for the data was at least 0.85, and in most cases >0.95. The phase-angle $\phi$ between the two signals was calculated as:

$$\phi = -(\phi_d - \phi_P), \qquad (2)$$

and attenuation was calculated by taking the tangent of $\phi$ as $Q^{-1} = \tan \phi$ (Lakes, 1998; Cooper, 2002).

## Machine damping

The inherent attenuation of the nanoindenter itself (hereafter referred to as machine damping) must be accounted for in estimating the material response of the tested materials. Machine damping was measured by performing the same forced oscillation method as for the tested materials, only without a specimen present (i.e., in air). Machine damping $Q_m^{-1}$ was calculated by taking the tangent of the phase angle $\phi_m$ between the applied load and the displacement of the indenter tip, following the same analysis for calculating $\phi$ for the tested materials. The load associated with indenting air with the nanoindenter is the required force needed to overcome the springs in the indenter's force head (Badt et al., 2024). The origin of the damping in the nanoindenter is likely due to eddy currents in the loading coil (G. Pharr, personal communication) that form due to the movement of the center plate within the three-plate capacitance gauge of the indenter head used to measure displacement (Badt et al., 2024). Air flowing into and out of the capacitance gauge (as the indenter tip moves up and down) may be another contribution to the observed damping in the nanoindenter. The machine damping has been added to figure 3 to compare to the measured viscoelastic material response of each material.

## Thermal drift during low-frequency oscillations

Nanoindentation tests can suffer from the inherent problem of thermal drift, where thermal fluctuations affect the spacing between the capacitors in the capacitance gauge (Verma et al., 2021; Badt et al., 2024), and thus the measured displacement (depth $h$) of the indenter tip. While over short timescales (≤300 s) these effects are often negligible, for longer periods they can lead to significant errors in indenter depth measurements (Thom et al., 2018; Verma et al., 2021, Badt et al., 2024). In our forced oscillation experiments, low-frequency oscillations (≤ 0.001 Hz) show complex depth-history responses that include a combination of both the material response to the cyclical loading and the effects of thermal drift (figure 2b). This complication makes the interpretation of the depth data difficult and impossible to fit to a simple sine function (e.g., figure 1d) without aggressive signal processing. To overcome the effects of thermal drift one may conduct indentation tests in an isolated environment with temperature control with fluctuations of < 0.5 °C/hour (Nohava et al., 2009). For the case of cyclic loading the effects of thermal drift can be eliminated when we compute the time derivative of the indentation depth $dh/dt$ over a short timescale where thermal drift is negligible. We calculated $dh/dt$ over a 1-s timescale (comparable to the sampling frequency in long-period oscillation tests with $T \geq 1000$ s) for the entire test (figure 2c). Since the indentation depth $h$ with time in cyclical loading is expected to follow a sine function as in equation (1), $dh/dt$ will follow a cosine function in the form of:

$$dh/dt = 2\pi f a_{est} \cos(2\pi f t + \phi_{est}) + c_{est}, \qquad (3)$$

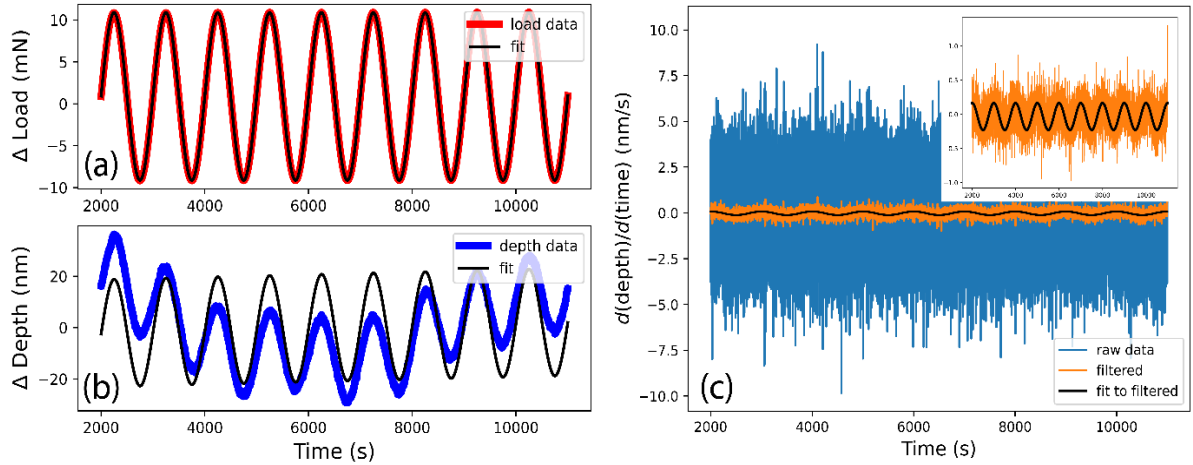

Figure 2. Forced oscillation test on San Carlos olivine with a loading frequency of 0.001 Hz and with mean load removed from plot (a) and the subsequent change in indenter depth showing the same oscillation frequency. The depth signal in (b) displays a complex behavior which includes changes in depth over time periods longer than the oscillation period (1000 s). The time derivative of the indenter depth with time is given in (c), where the light blue represents the computed raw data, the orange curve is the filtered signal, using a lowpass filter around the frequency of 0.001 Hz, and the black curve is a fit to the filtered signal. The inset in (c) is a blowup of the filtered signal and the fit over the same time range with the raw data (blue curve) removed for better visibility.

where $c_{est} = 0$ is the estimated mean for the cosine function. By fitting a function of the form of equation (3) to computed $dh/dt$ data one can find the phase of the depth signal $\phi_{est}$ without the effects of thermal drift (figure 2c). The raw $dh/dt$ data are filtered with a lowpass second-order Butterworth filter around the oscillation frequency $f$ (figure 2c) and the filtered signal is then fitted by equation (3) to find the phase lag $\phi_{est}$ of $dh/dt$. The phase lag between the load and the depth signals is then calculated as in equation (2), where $\phi_{est} = \phi_d$. We test the reproducibility of $Q^{-1}$ calculations based on fitting the depth data and fitting the time derivative of the depth data in figure S4 in the supporting information (SI) for tests on PMMA. The mean error between the phase lag calculated from indenter depth $\phi_h$ and the phase lag calculated from the depth-time derivative $\phi_{dh/dt}$ (computed as $(\phi_h - \phi_{dh/dt})/\phi_h$ ) was found to be 6.4% over the full range of frequencies tested (figure S4 in SI).

## Results

The results from all experiments are presented in figure 2 and compared to attenuation data from previous studies, where available (figure 3a, b). The machine damping $Q_m^{-1}$ results are plotted in comparison to the tested materials in figure 3 and show an increase in $Q_m^{-1}$ with oscillation frequency. Attenuation data for PMMA at room temperature (Lakes, 1998) were compared to attenuation data from our nanoindentation tests (figure 3a). The two data sets are in excellent agreement over the tested frequency range of 0.001-0.0333 Hz (figure 2a). At frequencies of 0.033 – 0.333 Hz, $Q^{-1}$ from nanoindentation plots systematically higher than the reported data from Lakes (1998). This trend coincides with an increase in machine damping (figure 3a), which suggests that the attenuation from the nanoindenter test is in reality lower and should plot closer to the reported data from Lakes (1998). At $f > 0.333$ Hz attenuation obtained from the naonindenter is too close to the machine damping to be of any significance. Results from forced oscillation tests on indium by nanoindentation are also shown in figure 3a and compared to a calculated attenuation spectrum (based on creep tests) by Ledbetter et al., (1996). Indium is a high-damping metal at room temperature due to the high homologous temperature $T_H$ = 0.69. The mechanism that controls attenuation at room temperature in indium is dislocation interactions and dislocation-point defect interactions within the specimen (Sapozhnikov et al., 2010). We conducted two sets of tests with the nanoindenter at loads of 2 mN (figure 3a - cyan triangles) and 5 mN (figure 3a – cyan circles). At these loads the contact area between the diamond tip and the indium specimen ($A_c$) is significantly different, as determined from the relationship between the nanoindentation hardness ($H$) and load (figure S3), where $H = P/A_c$ (e.g., Oliver and Pharr, 1992). The ratio of contact area at load $P$ = 2 mN to that under a load $P$ = 5 mN is $\frac{A_c^{P=2}}{A_c^{P=5}} = \frac{\frac{2}{0.045}}{\frac{5}{0.037}} \approx 0.33$, meaning that the contact area under the low load of 2 mN is about one third of the contact area under 5 mN, and as a result more dislocations are expected to affect the mechanical response of the material under the higher load of 5 mN. The attenuation data from the 5 mN test plot significantly higher than the data obtained by the 2-mN load tests with the nanoindenter (figure 3a). The data from the low load test with the nanoindenter compare well to a calculated attenuation spectrum from Ledbetter et al. (1996) for indium at room temperature. In addition, both data sets from the nanoindenter display peak attenuation around a frequency of ~0.003 Hz as show by the Ledbetter et al. (1996) data, giving further validation of our method.

Attenuation of single-crystal San Carlos olivine is presented in figure 3b and is compared to machine damping and to data obtained in D-DIA experiments on San Carlos olivine at room temperature at stress amplitudes of 1000 MPa (D. Hein, personal communication, figure S1 in SI). Olivine attenuation at room temperature is measurable with the nanoindenter at frequencies < 0.0033 Hz ($T$ > 300-s) where $Q^{-1}$ ~ 0.09-0.3 (figure 3b), an attenuation response similar to that obtained in the D-DIA experiments at $f$ = 0.0033 Hz (figure 3b). The attenuation data obtained in the nanoindenter represent multiple experiments sampled at different locations in the crystal with a load of 100 mN (corresponding to a stress of ~ 9 GPa and oscillation amplitude of 675 MPa). The different testing locations in the crystal may explain the spread of these data, due to an inhomogeneous distribution in the crystal of elements (e.g., dislocations) that affect attenuation. Another contribution may derive from the chosen background load of the forced oscillation. The attenuation data for indium from the nanoindenter suggests that at the higher load of 5 mN (and thus high contact area), the variability of the data for a certain frequency is lower compared to that in the 2-mN load tests (figure 3a). We chose a load of 100 mN in the olivine tests to prevent the occurrence of fractures during force oscillations. Testing olivine at higher homologous temperatures ($T_H$ > 0.19) will allow higher loads, and thus larger contact area, and will perhaps have less spread in the attenuation results at a given frequency.

Figure 3c presents attenuation results for a single crystal of quartz. These experiments were conducted at loads of 100 and 150 mN, which prevented fracturing of the sample. Attenuation was successfully measured in quartz for only the longest oscillation periods tested in this study, $3\times10^3$ and $1\times10^4$-s ($f \leq 3.33 \times 10^{-4}$ Hz). At higher frequencies, attenuation of quartz is slightly higher or at about the level of machine damping and so these measurements are questionable due to their high noise level.

Halite exhibited $Q^{-1}$ ~ 0.3 over the tested frequency range of 0.001 – 0.1 Hz, well above the machine damping level (figure 3d). At the highest tested frequency of 1 Hz, halite attenuation is only slightly higher than the machine damping. Attenuation in halite (and other rock salt formations) is typically tested at high frequencies (MHz), and these studies find that $Q^{-1}$ for compressional waves is 0.022 (Sears and Bonner, 1981), nearly an order of magnitude lower than what we measure with forced oscillations in the nanoindenter.

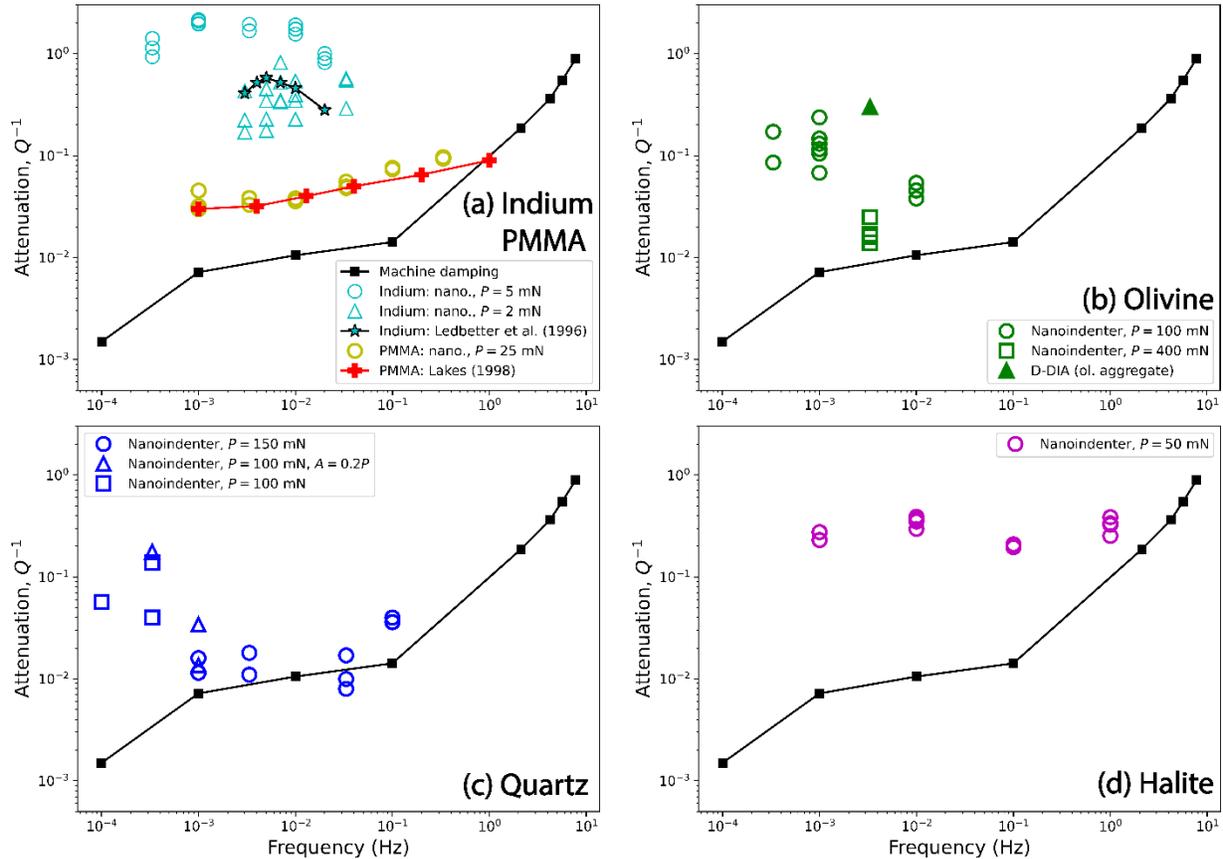

*Figure 3. Attenuation versus frequency results for commercial PMMA (a), San Carlos olivine (b), quartz (c) and halite (d). The machine damping results are presented in each individual plot by the black square-solid curve. The loads P for each material are given in the legends with oscillation amplitudes of $A = 0.1P$ (unless stated otherwise). Data from D-DIA experiments in (b) were provided by D. Hein (personal communication), where the oscillation amplitude was 700 MPa.*

## Discussion

We present a new method for measuring the viscoelastic response of materials by adapting the forced oscillation technique to the nanoindenter. This method is experimentally simple to set up, allowing for multiple tests to be performed on the same specimen and robust as shown by reproducing similar $Q^{-1}$ values to those measured by other apparatuses across a wide frequency range (figure 2). The uniqueness of our method allows us to probe microphysical mechanisms that control attenuation, which in single crystals are expected to be due to dislocation or point defect mechanisms (Karato and Spetzler, 1990). In contrast, comparatively larger-scale experiments, such as those conducted in the gas apparatus (e.g., Jackson, 1993), may involve activation of multiple deformation mechanisms (e.g., grain boundary sliding, subgrain boundary sliding, dislocation motion, point defects) which can make interpretation of results challenging.

Inducing anelasticity and plasticity in hard materials with high melting temperatures (> 1300 °C) at room temperature conditions is non-trivial (Frost and Ashby, 1982). However, forced oscillations in the nanoindenter activated attenuation mechanisms in quartz and olivine at low frequencies $\leq 10^{-3}$ Hz. In contrast, for comparatively soft materials like halite, indium and PMMA with indentation hardness $H < 1$ GPa (figure S3 in SI), attenuation was activated over a broad frequency spectrum of $10^{-3}$-1 Hz. Halite has a low yield strength of ~30 MPa at room temperature (Robertson et al., 1958) and its time-dependent deformation mechanism has been shown to be due to dislocation motion (Fokker and Kenter, 1994; Thom and Goldsby, 2019). PMMA, while not a crystalline material, has a very low glass transition temperature of $T_g$ = 106 °C (Hajduk et al., 2021), which, when coupled with a low yield strength of ~100 MPa (Jin et al., 2016) can explain the attenuation behavior of PMMA at room temperature across all tested oscillation frequencies. Attenuation in indium has been shown to be dependent on dislocation interactions at room temperature in previous studies (e.g., Sapozhnikov et al., 2010) which may be evident in our experiments on indium with the nanoidenter at low ($P$ = 2 mN) and high ($P$ = 5 mN) loads. The harder materials studied here, olivine and quartz ($H \sim 14$ GPa, figure S3 in SI), both tested at very low homologous temperatures of <0.19, exhibited measurable attenuation only at low frequencies of $f \leq 10^{-3}$ Hz. At these conditions plasticity is expected to control crystal deformation (Frost and Ashby, 1982) but is not measurable with the nanoindenter at high strain rates, $f >$ 0.001 Hz for quartz and $f >$ 0.01 Hz in olivine due to machine damping. In addition, the relative spread in the attenuation results at low frequencies in quartz and olivine suggest that the plastic response of the material needs to be tested at multiple locations within the same crystal to obtain a bulk viscoelastic response representative of the whole crystal.

At higher temperatures dislocation motion is expected to contribute to the viscoelastic behavior in the solid as obstacles, such as jogs, are more easily overcome (Karato, 2008). This opens the possibility of performing forced oscillations with the nanoindenter at elevated temperatures with a dedicated heating stage (e.g., Schuh et al., 2005) to probe the micromechanics of olivine and quartz attenuation over a wider range of frequencies. Obtaining high-temperature measurements of attenuation in single crystals of olivine and quartz is a crucial step towards better understanding of the physical processes that control large-scale phenomena such as mantle convection in the Earth. Quantifying the creep and anelastic behavior of minerals at high temperatures and pressures will provide better constraints on geophysical models of Earth and other planetary systems. Forced oscillation tests in the nanoindenter have been shown here to be a potentially powerful tool to achieve these measurements with both accuracy and simplicity of experimental design.

## Conclusions

A new method for measuring attenuation through nanoindentation shows promise in quantifying the attenuation spectrum of geomaterials. The method has been shown to be successful in reproducing attenuation values obtained from other apparatuses in PMMA and indium. The method itself is experimentally simple to set up and multiple indents can be performed on the same specimen with high spatial resolution. At room temperature conditions attenuation of soft materials can be performed over a broad frequency range of 0.001-1 Hz, whereas for harder materials with high melting temperatures, such as quartz and olivine, attenuation can be measured only at low frequencies of <0.001 Hz. Attenuation in quartz and olivine are expected to be measurable at high temperatures and over a broad range of frequencies in the nanoindenter, when nanoindentation is combined with a sample heating stage.

**Data availability statement**

The data are available online at Zenodo (Badt, 2024a, 2024b).